\documentclass[12pt,preprint]{aastex}

\def\aa#1#2#3#4#5{\bibitem[#1]{#2}#3, {A\&A}, {#4}, #5}
\def\aasup#1#2#3#4#5{\bibitem[#1]{#2}#3, {A\&AS}, {#4}, #5}
\def\aj#1#2#3#4#5{\bibitem[#1]{#2}#3, {AJ}, {#4}, #5}
\def\apj#1#2#3#4#5{\bibitem[#1]{#2}#3, {Ap. J.}, {#4}, #5}

\def\apjsup#1#2#3#4#5{\bibitem[#1]{#2}#3, ApJS, #4, #5}
\def\araa#1#2#3#4#5{\bibitem[#1]{#2}#3, ARA\&A, #4, #5}

\def\mnras#1#2#3#4#5{\bibitem[#1]{#2}#3, {M.N.R.A.S.}, {#4}, #5}
\def\nature#1#2#3#4#5{\bibitem[#1]{#2}#3, {Nature}, {#4}, #5}
\def\pasj#1#2#3#4#5{\bibitem[#1]{#2}#3, PASJ, #4, #5}

\begin{document}

\title{Outflow 20 -- 2000 AU from a High-Mass Protostar in W51-IRS2}

\author{J. A.  Eisner \altaffilmark{1,2,3},
L. J. Greenhill \altaffilmark{1},
J. R. Herrnstein \altaffilmark{2,4}, J. M. Moran \altaffilmark{1},
\& K. M. Menten \altaffilmark{1,5}
\email{jae@astro.caltech.edu}}

\altaffiltext{1} {Harvard-Smithsonian Center for Astrophysics,
60 Garden Street, Cambridge, MA 02138}
\altaffiltext{2} {National Radio Astronomy Observatory, P.O. Box 0,
Socorro, NM 87801}
\altaffiltext{3} {Current Address: Palomar Observatory 105-24,
California Institute of Technology, Pasadena, CA 91125}
\altaffiltext{4} {Current Address:  Renaissance Technologies
Corporation, 600 Route 25A, East Setauket, NY 11733}
\altaffiltext{5} {Current Address: 
Max-Planck-Institut f\"{u}r Radioastronomie, 
Auf dem H\"{u}gel 69, D-53121 Bonn, Germany}


%

\begin{abstract}
We present the results of the first high angular resolution 
observations of SiO maser emission towards the star forming region
W51-IRS2 made
with the Very Large Array (VLA) and Very Long Baseline Array (VLBA).
Our images of the H$_2$O maser emission in 
W51-IRS2 reveal two maser complexes
bracketing the SiO maser source.  
One of these H$_2$O maser complexes appears to trace
a bow shock whose opening angle is consistent with the opening
angle observed in the distribution of SiO maser emission.
A comparison of our H$_2$O maser image with an image constructed 
from data acquired
19 years earlier clearly shows the persistence and motion of
this bow shock.  The proper motions correspond to an outflow velocity
of 80 km s$^{-1}$, which is consistent with the data of 19 years ago
(that spanned 2 years).
We have discovered a two-armed linear structure in the SiO maser emission
on scales of $\sim 25$ AU,
and we find  a velocity gradient on the order of $0.1$ km s$^{-1}$ AU$^{-1}$
along the arms.
We propose that the SiO
maser source traces the limbs of an accelerating bipolar outflow
close to an obscured protostar.  We estimate that the outflow
makes an angle of $< 20^{\circ}$ with respect to the plane
of the sky.  Our measurement of the acceleration is consistent
with a reported drift in the line-of-sight velocity
of the W51 SiO maser source.
\end{abstract}

\section{Introduction}
Since the discovery of the first bipolar outflows in star-forming
regions \citep{SNELL+80}, significant progress has been made 
in understanding the large-scale characteristics of such outflows
\citep[{e.g.,}][]{BACHILLER96}.  However, the large columns of
gas and dust that obscure massive protostars hinder traditional optical
or infra-red observations of at least 
the inner $\sim 100$ AU of these outflows,
where the exciting protostars reside. This has made it difficult to
obtain sufficient data to understand well the process of high-mass star 
formation. 
Radio frequency maser emission, 
which traces velocity coherent clumps of gas entrained
in large-scale bulk mass motions,
is unattenuated by the neutral gas and dust around massive protostars.
Moreover, because masers are compact, high-brightness sources,
high angular resolution radio interferometry can be used to probe
the structure and
kinematics of these outflows on angular scales of $\lesssim 1$ AU
for many Galactic sources \citep[{e.g.,}][]{GREENHILL+98, PATEL+00}.

Many protostellar outflows
exhibit H$_2$O maser emission at 22 GHz 
\citep[{e.g.,}][]{HENNING+92,FELLI+92} which traces shocks in 
dust-laden gas close to the exciting protostars \citep{ELITZUR92}.  
SiO masers, which are a common feature of
late-type stars, are known to occur 
in only three regions of star formation: W51-IRS2, Sgr-B2 MD5, 
and Orion-KL \citep{HASEGAWA+86, SB74}.
Maser action in vibrationally excited states of SiO at 43 GHz
requires higher temperatures 
($>10^3$ K) and is more closely associated with
exciting sources than is H$_2$O maser emission \citep{ELITZUR92}. 
The only non-stellar SiO maser
source that has been well studied is the one in 
Orion-KL, where the maser emission traces an outflow within $\sim 100$
AU of an obscured massive protostar \citep{GREENHILL+98, DLP99}.

W51-IRS2 is an embedded infrared source ($L_{\rm tot} \sim 
2.8 \times 10^6 \: L_{\odot}$; Erickson \& Tokunaga 1980) in the well-known
high-mass star forming
region W51. We adopt a distance of 7 kpc to W51 based on maser
proper motion studies \citep{GENZEL+81}.
The region around IRS2 contains an edge-brightened 
cometary HII region called W51d 
\citep{MARTIN72, WC89, GJW93} that is associated with a 2.2 $\mu$m
point source \citep{GW94} and a peak in 12 $\mu$m
emission \citep{OKAMOTO+01}.  An ultra-compact HII region called W51d2 
\citep{MARTIN72} is associated with NH$_3$ and methanol 
masers \citep{GJW93,
MCB01}
as well as a feature in the 12 $\mu$m continuum \citep{OKAMOTO+01}.
W51-IRS2 also contains an H$_2$O maser complex called W51 North, for which
detailed distributions and proper motions have been observed
\citep{SCHNEPS+81}, OH maser emission \citep{GM87},  
and an unresolved SiO
maser source \citep{HASEGAWA+86, UKITA+87, MORITA+92}.  
The strongest H$_2$O masers, the OH masers, and the SiO masers,
are found in a compact region of W51 North
termed the ``Dominant Center'' by Schneps et al. (1981).  Although
no infrared or radio continuum sources have been
detected  within $\sim 2''$ of it \citep{GJW93, WC89, OKAMOTO+01}, 
the Dominant Center does coincide with thermal emission from several
molecular species including NH$_3$(1,1), NH$_3$(2,2), NH$_3$(3,3), 
CS, and CH$_3$CN \citep{ZH97,ZHO98}.  

The spectrum of the H$_2$O maser emission covers $V_{\rm LSR} = -30$
to 130 km s$^{-1}$, and is peaked around 60 km s$^{-1}$ 
\citep{SCHNEPS+81}.  The SiO emission is peaked around 45 km s$^{-1}$
and covers $V_{\rm LSR} = 40$ to 50 km s$^{-1}$, lying within
the velocity range of the H$_2$O maser emission \citep{MORITA+92}.
The spectra for various molecular species
are peaked around $V_{\rm LSR} \sim 60$ km s$^{-1}$,
with typical linewidths of $\sim 20$ km s$^{-1}$ \citep{ZH97,ZHO98}.

In this paper, we present the first high angular resolution observations
of the SiO maser emission in W51-IRS2.  In \S \ref{sec:obs}
and \S \ref{sec:results}, we describe 
interferometric observations of the SiO and H$_2$O maser emission.
We interpret our results in the context of
a bipolar outflow model in \S \ref{sec:dc-disc}.
We also show that previous measurements of the proper motions
of H$_2$O masers and the velocity drifts of SiO masers
provide support for our model.

\section{Observations and Data Reduction}
\label{sec:obs}

\subsection{VLA Observations and Data Reduction}
\label{sec:obs-vla}
We observed the $^{28}$SiO $v=2$, $J=1 \rightarrow 0$ maser line
at 42.820539 GHz, the $6_{1,6} \rightarrow 5_{2,3}$ H$_2$O maser
line at 22.235080 GHz, and continuum emission at 22 and 43 GHz
in W51-IRS2.  On 1998 March 30, we used the most
extended configuration of the VLA of the NRAO,\footnote{The 
National Radio Astronomy Observatory is a facility of the National Science 
Foundation operated under cooperative agreement by Associated
Universities, Inc.} obtaining an angular resolution
of $\sim 40$ milliarcseconds (at 43 GHz) in 6 hours of on-source integration.
We observed in two bands with two sub-arrays, one at 22 and one at 43 GHz.
We tuned one 6 MHz band in
each sub-array to include a strong maser line, and a second 25 MHz band to
line-free continuum.  The channel spacing for the line observations
was $\sim 0.7$ km s$^{-1}$ at
43 GHz and $\sim 1.3$ km s$^{-1}$ at 22 GHz.  The intent of these
observations was to establish accurate relative positions among sources at
each frequency.

A second observing run, on 1998 July 25, was intended 
primarily to establish registration of the 22 GHz and 43 GHz images.
We observed the SiO and strong H$_2$O maser lines 
with the second-most extended configuration of the VLA, and obtained
an angular resolution of $\sim 0\rlap{.}''15$ (at 43 GHz).  
Observing in two bands with
two sub-arrays (as above), we ``fast-switched'' between W51-IRS2 and
a calibrator (J1925+2106)
with an 80s/80s duty cycle, achieving 40
minutes on-source.  We determined the positions
of the SiO and strong H$_2$O maser features with respect to the
calibrators to better than 50 mas.
We also observed the 22 GHz line-free
continuum for $\sim 2$ hours with two 25 MHz bands 
(in dual-polarization mode), with the goal of searching for
continuum emission associated with the Dominant Center that was not
detected on March 30 or by others \citep{GJW93, WC89}.  

We made images from the data of both observing runs with
standard techniques using the NRAO AIPS package 
\citep[{e.g.,}][]{RUPEN99}.  We calibrated the flux scale of the data
with observations of 3C286 (which had a flux of 
2.52 Jy at 22 GHz and 1.47 Jy at 43 GHz), 
and determined the bandpass response
using NRAO530.  For the March 30 observations, we 
self-calibrated the interferometric data using 
strong SiO and H$_2$O maser emission lines at $V_{\rm LSR} = 50$ 
and 65 km s$^{-1}$,
respectively.  Because the 22 GHz primary beam included W51 North
and W51 Main, we mapped both regions together. 
The gain solutions determined from the strong maser
lines were then applied to the rest of the maser emission, as
well as to the line-free continuum \citep[{e.g.,}][]{MR97}.  
By calibrating the continuum in this way, we achieve relative
astrometry between the maser and continuum emission that is largely
free of systematic effects caused by the troposphere.  The
registrations of the 22 GHz continuum to the H$_2$O maser emission
and of the 43 GHz continuum to the SiO maser emission are
accurate to $\ll 50$ mas, and are largely noise-limited.  
The registration between the 22 and
43 GHz images is uncertain by $\sim 0\rlap{.}''1$.
The formal astrometric
position for the SiO maser (the strongest feature at 50 km s$^{-1}$ on July
25, 1998) is: 
$(\alpha, \delta)_{\rm J2000}=$ 
($19^{\rm h}23^{\rm m}40\rlap{$^{\rm s}$}.055 \pm 0\rlap{$^{\rm s}$}.003$, 
$14^{\circ}31'5\rlap{.}''59\pm 0\rlap{.}''05$).
We adopt these coordinates as the reference for all relative images presented
in this paper.

We deconvolved the point source response from each image and fitted a
2-D elliptical Gaussian to each identified maser spot. (We 
define a maser ``spot'' as emission occurring
in a given velocity channel.  A maser ``feature'' refers to
spots clustered on the sky in $\lesssim 1$ beamwidth, which lie in a 
contiguous range of channels.  Maser features often correspond
to physically distinct clumps of gas and separate Doppler components
in spectra.)  The
associated statistical uncertainty in the fitted position of each centroid
is $0.5 \times \theta_{\rm beam} / {\rm SNR}$, 
where SNR is the signal-to-noise ratio, and
$\theta_{\rm beam}$ is the synthesized half-power beamwidth 
\citep{RM88}. For these observations, the synthesized beam was 
$0\rlap{.}''040 \times 0\rlap{.}''039$ with PA$= 1^{\circ}$ at 43 GHz 
and $0\rlap{.}''11 \times 0\rlap{.}''09$ with PA$ = -70^{\circ}$ at 22 GHz.
We note that the statistical uncertainty applies to
largely unresolved or point source emission.  It does not apply if there
is structure on scales comparable to the beam size, in which
case the fitted 
centroid corresponds to a flux-weighted mean of all the emission
regions within the beam.

\subsection{VLBA Observations and Data Reduction}
\label{sec:vlba-dr}
On 1994 July 6, we observed the $^{28}$SiO $v=2$, $J=1 \rightarrow 0$ 
maser with the
VLBA of the NRAO, integrating approximately 3 hours on-source.  At the time
of these observations, the array consisted of the Pie
Town, Kitt Peak, Los Alamos, Fort Davis, Owens Valley, North Liberty,
and Brewster antennas, which afforded an angular resolution of $\sim 0.6$ mas.
We recorded an 8 MHz band and correlated 256 channels
(with the correlator in Socorro, NM), which provided a velocity
resolution of $\sim 0.2$ km s$^{-1}$.

We calibrated these data with standard techniques of spectral-line VLBI
\citep[{e.g.,}][]{WALKER99, REID99}.  
We achieved an amplitude calibration accurate to
approximately 50\% using measured system temperatures and {\it a priori}
antenna gains.  Global fringe fitting to continuum
calibrators (3C273B, 3C345, 3C454.3, J1800+7828, J1740+5211)
provided estimates of 
instrumental delay and fringe rate.  We self-calibrated using the
peak maser emission at $V_{\rm LSR} = 43$ km s$^{-1}$, and applied 
the solutions to the rest
of the maser data.  As was done for the VLA data, maser spot positions
and velocities were determined by fitting 2-D elliptical Gaussian models
to maser spots in deconvolved images.
The synthesized beam was $0.89 \times 0.52$ mas at PA$ = -38^{\circ}$.

\section{Results}
\label{sec:results}

\subsection{H$_2$O Masers in W51-IRS2}
\label{sec:h2o}
\begin{figure}
\plotone{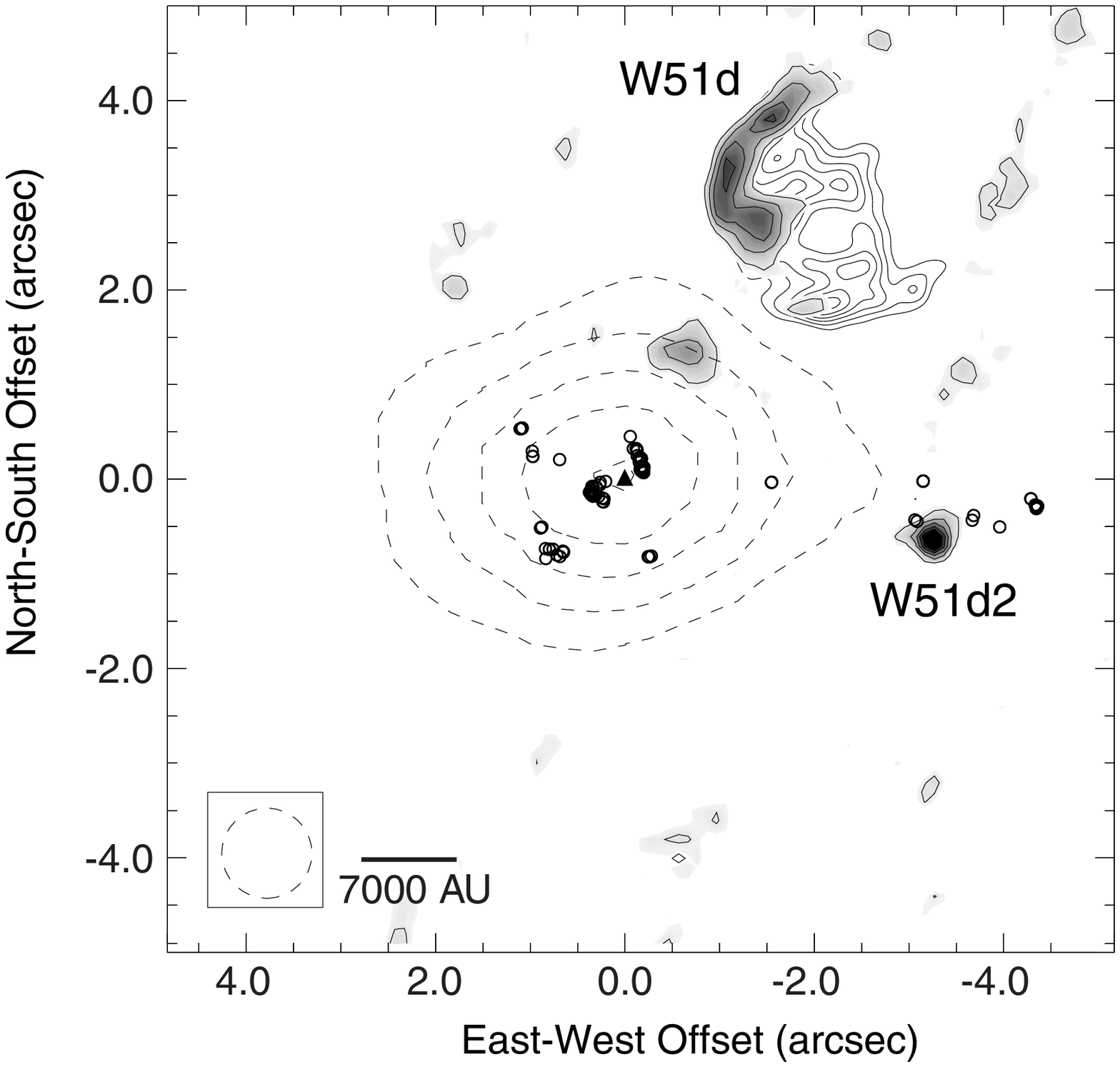}
\caption{The W51-IRS2 star forming region, mapped with
the VLA.  
The coordinates are offset from the position of the SiO maser source
(indicated by the black triangle) at
$(\alpha, \delta)_{\rm J2000}=$ 
($19^{\rm h}23^{\rm m}40\rlap{$^{\rm s}$}.055 \pm 0\rlap{$^{\rm s}$}.004$, 
$14^{\circ}31'5\rlap{.}''59\pm 0\rlap{.}''07$).
The continuous contours represent
continuum emission at 22 GHz observed with $0\rlap{.}''24$
resolution (filled contours starting at 8 mJy, separated by
4 mJy), and observed at $0\rlap{.}''47$ resolution (unfilled contours 
starting at 8 mJy, separated by 1.7 mJy).  The rms noise levels are
1.0 mJy and 1.7 mJy, respectively.  The dashed contours represent thermal
emission from the $(J,K)=(1,1)$ transition 
in NH$_3$ observed with $\sim 1''$ resolution \citep{ZH97}.
The open circles indicate H$_2$O maser emission (at 22 GHz), and the
triangle represents the position of the SiO maser source
(at 43 GHz). The previously identified sources W51d and W51d2
are labeled.  The relative astrometry of SiO masers, H$_2$O masers, and
22 GHz continuum sources is accurate to $\lesssim 0\rlap{.}''1$.
\label{fig:w51-big}}
\end{figure}

\begin{figure}
\plotone{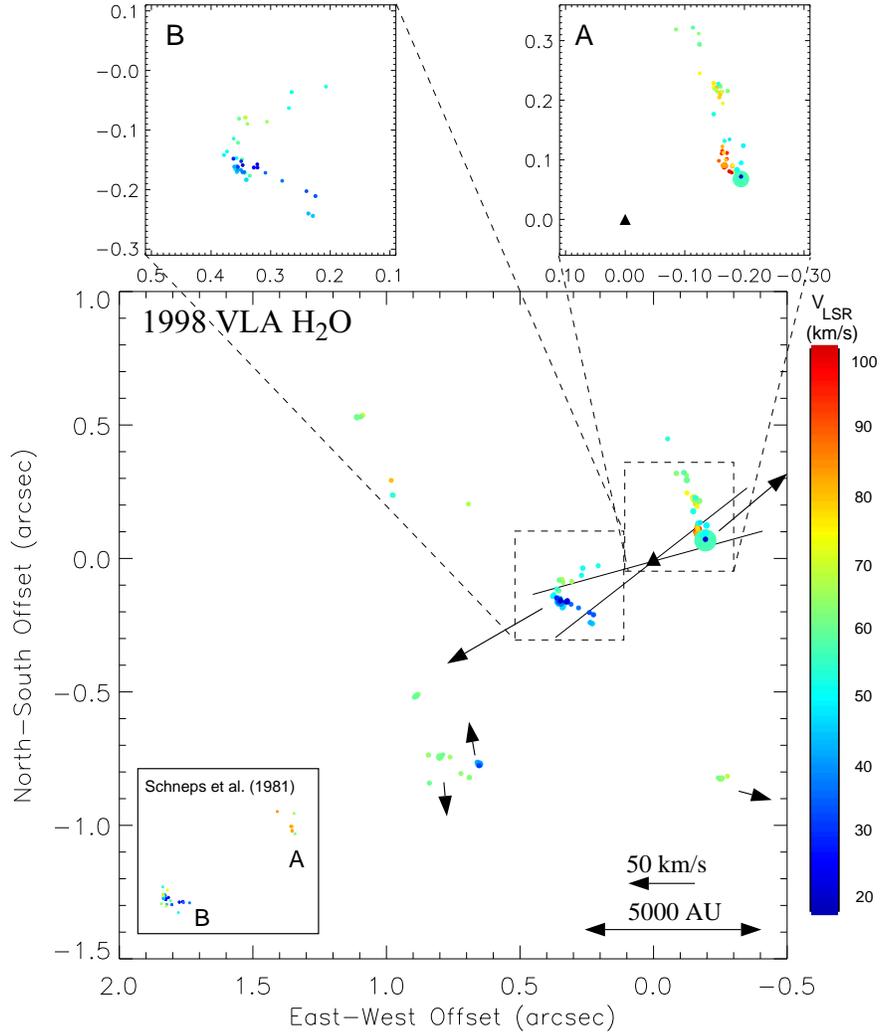}
\caption{The distribution of H$_2$O maser emission in the Dominant Center 
Region, observed with the VLA on 1998 March 30.
The coordinates are offset from the VLA astrometric 
position of the SiO maser source
(indicated by the black triangle) at
$(\alpha, \delta)_{\rm J2000}=$ 
($19^{\rm h}23^{\rm m}40\rlap{$^{\rm s}$}.055$,
$14^{\circ}31'5\rlap{.}''59$).
The circles represent H$_2$O
maser spots, where color indicates line-of-sight velocity and area is
proportional to flux density.  The uncertainties in the 
relative fitted positions 
of the maser spots are smaller than the sizes of the plotted symbols.
The registration error of this map with respect to
the SiO reference position is $0\rlap{.}''07$ in each coordinate.
The arrows denote average measured
proper motions of the clusters of H$_2$O masers measured with
data from 1979 \citep{SCHNEPS+81}. 
The opening angle of the SiO maser emission seen in Figure 
\ref{fig:siomasrs-vlba} has been extrapolated to the scales of the 
H$_2$O maser emission (solid lines).
Blow-ups of two clusters of masers, denoted ${\cal A}$ and ${\cal B}$
are shown at the top.  The inset shows the maser distributions of
${\cal A}$ and ${\cal B}$ in 1979.
\label{fig:dc}}
\end{figure}

We imaged seven clusters of H$_2$O maser emission in 
W51-IRS2 (with an rms sensitivity of $\sim 5$ mJy) , 
and we covered velocities from $V_{\rm LSR} = 20$ 
to 100 km s$^{-1}$ (Figures \ref{fig:w51-big} and \ref{fig:spec}).
We note that our velocity coverage does not include the
entire spectrum, which extends from $V_{\rm LSR} \sim -30$
to 130 km s$^{-1}$ \citep{SCHNEPS+81}.  One cluster of maser spots is
associated with, but offset from, the ultracompact HII region W51d2, and
five other clusters of maser spots are coincident with a dense molecular core
traced by thermal emission from NH$_3$(1,1) and other
molecular tracers (Figure \ref{fig:w51-big}).  A seventh cluster 
(containing only two maser spots) lies 
between W51d2 and the NH$_3$ peak.
With the exception of the cluster at the northeast corner of the
Dominant Center region and the cluster between W51d2 and the NH$_3$ emission
peak, all were previously observed by Schneps
et al. (1981).  A comparison of our image with Schneps et al. (1981) 
reveals a remarkable stability in the structure of the H$_2$O maser
source over a 19 year baseline (Figure \ref{fig:dc}).
In the remainder of this discussion, we 
will focus on the H$_2$O maser clusters labeled ${\cal A}$ and 
${\cal B}$ in Figure \ref{fig:dc}.

The strongest water maser emission in the Dominant Center
${\cal A}$, the SiO maser source, and the
apparent vertex of the central cluster of H$_2$O maser spots
${\cal B}$ are approximately colinear
on the sky.  The position angle of this line is $\sim 108^{\circ}$, 
and the distance between
${\cal A}$ and ${\cal B}$ is $\sim 4270$ AU (assuming a distance of
$\sim 7$ kpc).  Schneps et al. (1981) measured proper motions of the
H$_2$O masers from 1977 to 1979 that indicate that ${\cal A}$ and
${\cal B}$ (denoted ``NW Cluster'' and ``Dominant Center Reference''
in their paper)
are moving away from each other in the plane of the sky at 
$v = 160 \pm 40$ km s$^{-1}$ along this position angle.
(We estimated the uncertainty in this
determination from the dispersion in the proper motions of 
five maser spots in the
Dominant Center Reference cluster from the data in Schneps et al.)
We confirm the proper motion by
comparing the separation of clusters ${\cal A}$ and ${\cal B}$ from
Schneps et al. (1981) to the separation measured by us 19 years later
(Figure \ref{fig:dc}).
We measure an increase of separation from $0\rlap{.}''52$
to $0\rlap{.}''61$ and find
$v \sim  150$ km s$^{-1}$, which
is consistent with Schneps et al. (1981).  Imai et al. (2001) 
present an independent estimate of the proper
motions based on recent short time-baseline VLBA observations,
and report a relative velocity between ${\cal A}$ and ${\cal B}$
of $\sim 200$ km s$^{-1}$ \citep{IMAI+01}.

In addition to confirming the ${\cal A}$-${\cal B}$ proper motion 
measured by Schneps et al. (1981),
our observations show
that the H$_2$O maser structures retained their shapes as they moved
apart.  Specifically, as shown in Figure \ref{fig:dc}, 
the structures of ${\cal A}$ and ${\cal B}$ have not changed significantly
over the 19 years. (Slight discrepancies between the
two images can be partially attributed to the fact that 
Schneps et al. (1981) covered velocities from -10 to 80 km s$^{-1}$,
while the observations presented here covered velocities from 20 to 100 
km s$^{-1}$.)  This is particularly noteworthy because the distribution
of H$_2$O masers in cluster ${\cal B}$ suggests that it traces the 
limb of a bow shock, and we actually see this bow shock moving 
persistently outward over a 19 year interval.

\subsection{SiO Maser Source}
\label{sec:sio}
\begin{figure}
\plotone{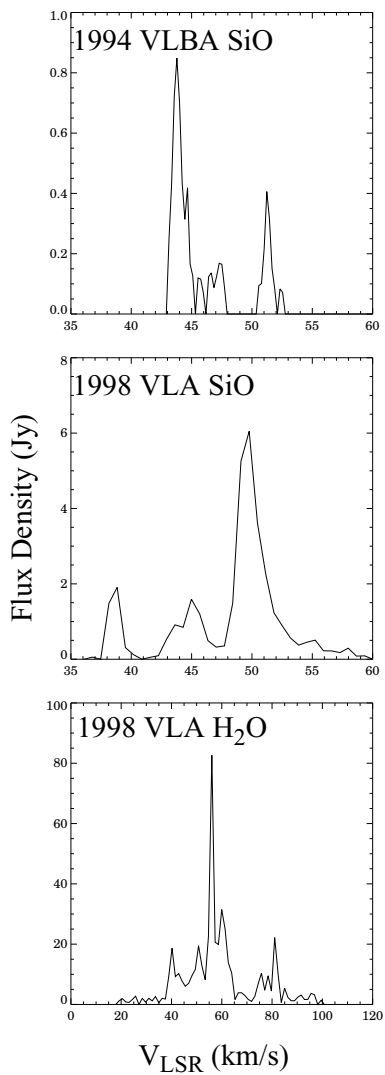}
\caption{(top) Spectrum of the SiO maser emission observed on 1994 July 6
with the VLBA, (middle) and on 1998 March 30 with the VLA.  Note
that the flux in the VLBA spectrum is much less than the VLA spectrum,
in part because the former is sensitive to only the very small scale emission.
(bottom) Band-limited spectrum of the H$_2$O maser 
emission observed on 1998 March 30 with the VLA.
Emission has been observed previously over a broader velocity range,
--30 to 130 km s$^{-1}$ \citep{SCHNEPS+81}.
\label{fig:spec}}
\end{figure}

The SiO maser emission lies in between the H$_2$O maser complexes ${\cal A}$ 
and ${\cal B}$ on the sky and is effectively coincident with the 
peak of thermal NH$_3$ emission (Figure \ref{fig:w51-big}).
The spectrum of the SiO maser emission covers velocities from 
$V_{\rm LSR} = 38$ to 60 km s$^{-1}$, which lie within the velocity
extent of the H$_2$O maser emission (Figure \ref{fig:spec}), but are
offset from the peak of thermal 
emission from NH$_3$ (and other molecular tracers)
at $\sim 60$ km s$^{-1}$ \citep{ZH95,ZHO98}.

\begin{figure}
\plotone{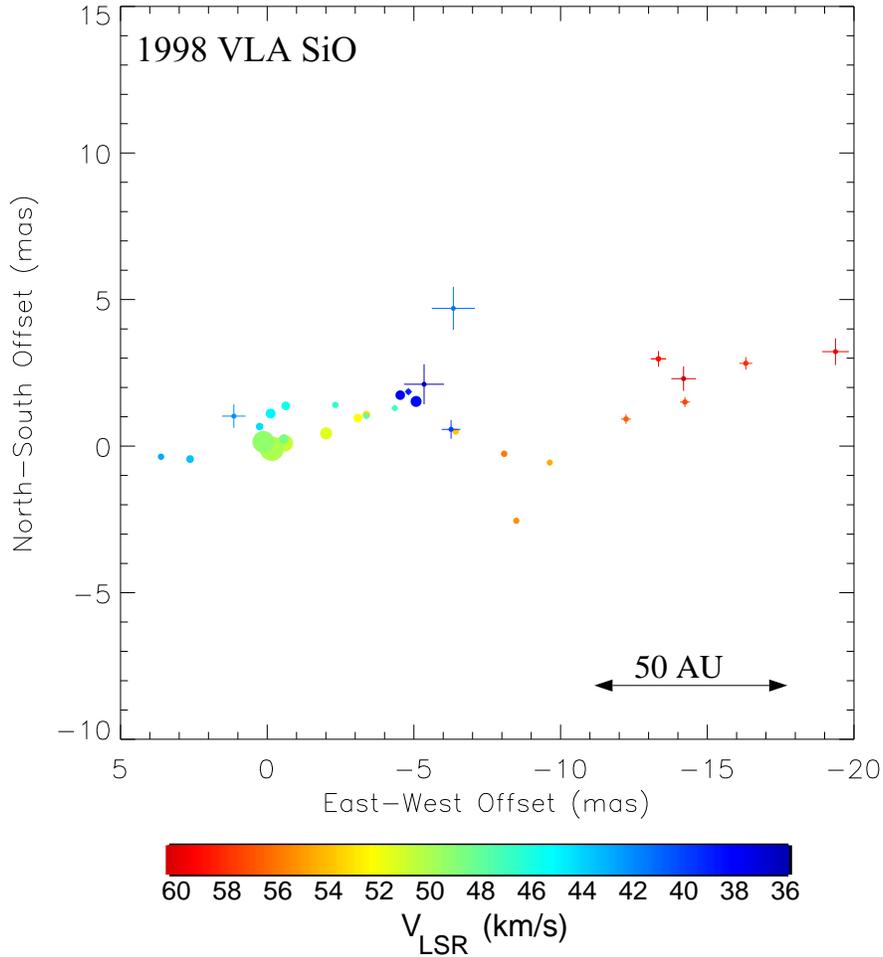}
\caption{The W51-IRS2 SiO maser source, mapped with the VLA at
40 mas resolution on 1998 March 30. The coordinates in this figure are
offset from $(\alpha, \delta)_{\rm J2000}=$ 
($19^{\rm h}23^{\rm m}40\rlap{$^{\rm s}$}.055$, 
$14^{\circ}31'5\rlap{.}''59$). These coordinates 
correspond to the black triangle plotted in Figures \ref{fig:w51-big}
and \ref{fig:dc}.  The absolute registration
of the image is accurate to 50 mas in each coordinate.
The circles represent individual maser
spots, where the color corresponds to line-of-sight velocity, and area is
proportional to flux density.  Error bars denote $1\sigma$ uncertainties
in relative position for the emission centroid at each velocity.
Note the narrower velocity range here
compared to Figure \ref{fig:dc}.
\label{fig:siomasrs}}
\end{figure}

We resolved the angular structure of the W51-IRS2 SiO maser source
with both the VLA and the VLBA data.  The source has a linear structure
with a position angle of $\sim 105^{\circ}$ (Figures \ref{fig:siomasrs} 
and \ref{fig:siomasrs-vlba}).  The SiO maser also shows an apparent
gradient in line-of-sight (LOS) velocity with position, with redder
spots to the west and bluer spots to the east.  From the data
shown in Figures 
\ref{fig:siomasrs} and \ref{fig:siomasrs-vlba}, 
we estimate this gradient to be $\sim 1$ km s$^{-1}$ mas$^{-1}$
($\sim 0.1$ km s$^{-1}$ AU$^{-1}$).
The SiO maser emission seems
to be distributed along a northern and a southern ``arm'', and the
LOS velocities of the two arms are offset by 5 to 10 km s$^{-1}$
(Figure \ref{fig:siomasrs-vlba}).

The VLA image is more sensitive to the large scale structure in the SiO
maser source but less accurate in pinpointing emission centers than the
VLBA.  Specifically, the rms sensitivities are 7 mJy for the VLA and 14 mJy
for the VLBA, while the angular resolutions are $\sim 40$ mas and 
$\sim 0.6$ mas,
respectively.  Thus, while the observed emission in the VLA image may be
a flux-weighted centroid of structures smaller than the beamsize, the
VLBA image reveals the positions of the compact core components.
The total flux in the VLBA spectrum is considerably less than that
of the VLA spectrum (Figure \ref{fig:spec}), which is probably due to
the different instrument resolutions and/or source variability.

\subsection{Continuum Emission} 
\label{sec:cont}
Figure \ref{fig:w51-big} shows the radio continuum emission in the vicinity
of W51-IRS2.  Detection of W51d and W51d2 is consistent with previous 
observations \citep{GJW93,WC89}.  
We did not detect any radio continuum emission in the Dominant Center
region, setting an upper limit of $\sim 0.1$ mJy
at 22 GHz \citep[as compared to the previous upper limit of $\sim 1$ mJy;][]
{GJW93}.  
The lack of radio continuum emission in a dense molecular core
traced by maser emission and thermal NH$_3$ emission indicates that
any compact optically thick
HII region in the Dominant Center is too small to be detected
({i.e.} $\lesssim 50$ AU at $10^4$ K), perhaps due to confinement by
accretion from the surrounding molecular cloud \citep[{e.g.,}][]{OSORIO+99}
or by outflow (or both).

\section{Discussion: Outflow in the Dominant Center Region}
\label{sec:dc-disc}
The positional coincidence and velocity overlap of the SiO and 
H$_2$O maser emission 
with a peak in the thermal NH$_3$ emission (Figures 
 \ref{fig:w51-big} and \ref{fig:spec}) probably signifies that the masers 
trace star forming activity in the denser portions of 
a molecular cloud core.  
The outflowing motions of the H$_2$O masers on scales of $\sim 4200$ AU
and the apparent bow shock structure ${\cal B}$ 
also support this conclusion.

We suggest that the SiO maser source in W51-IRS2 marks the position of a
massive protostar, because of excitation requirements of the $v=2, J=1 \rightarrow 0$ SiO maser emission
\citep[$n_{H_2} 
\sim 10^{10}$ cm$^{-3}$, $T_{\rm ex} \sim 3500$ K;][]{ELITZUR92}.
Thus, the absolute position of the
protostar is within 5 mas of our 
reference VLA SiO astrometric position,
$(\alpha,\delta)_{\rm J2000} =$ ($19^{\rm h}23^{\rm m}40
\rlap{$^{\rm s}$}.055 
\pm 0\rlap{$^{\rm s}$}.003$,
$14^{\circ}31'5\rlap{.}''59 \pm 0\rlap{.}''05$). 
We surmise that the exciting protostar is massive based
on an estimate of the mechanical luminosity of the H$_2$O maser
outflow:  assuming an H$_2$ density for 
the outflow of $\ge 10^6$ cm$^{-3}$ (realistic based on the
H$_2$O and OH maser emission in the vicinity; Elitzur 1992), 
an opening angle of $25^{\circ}$,
and flow velocity of $\sim 80$ km s$^{-1}$,
we find $L \ge 10$ $L_{\odot}$.

\begin{figure}
\plotone{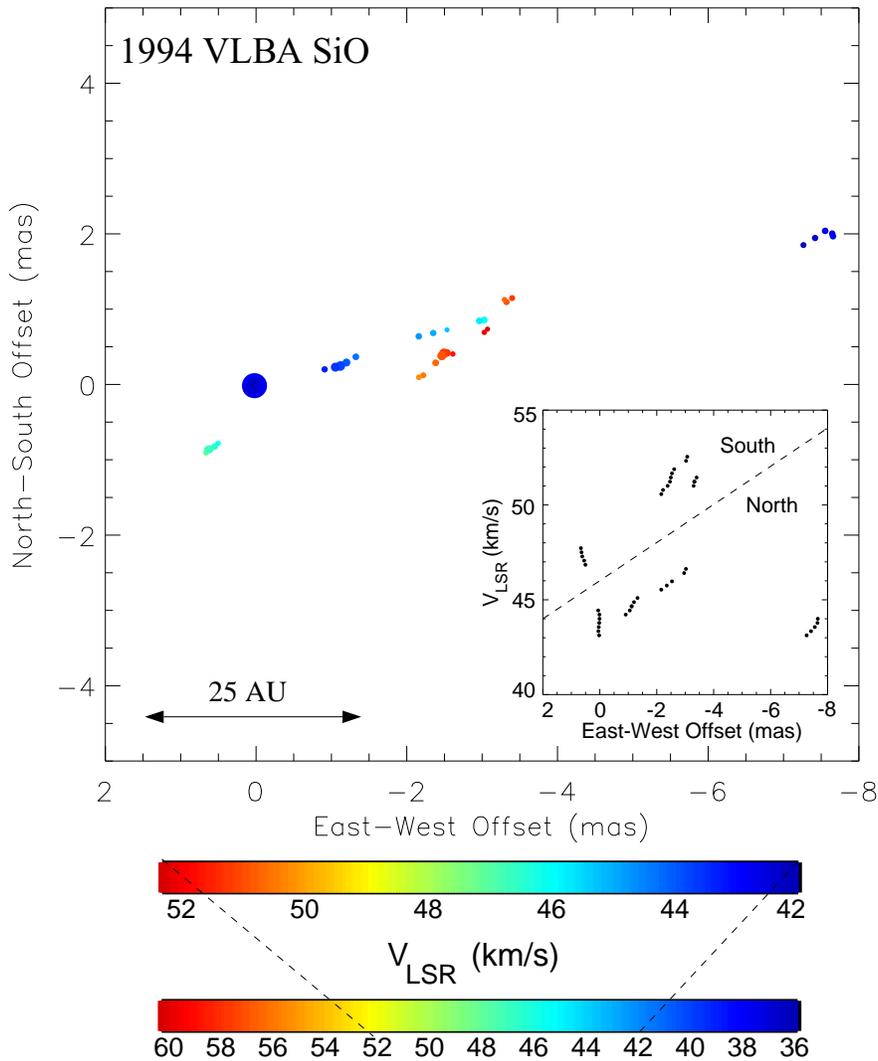}
\caption{The W51-IRS2 SiO maser source, mapped with the VLBA at 0.48
mas resolution, on 1994 July 6.  The coordinate offsets are
with respect to the strongest emission at the time of observation.
No VLBA astrometric data were available to register this image
with the one in Figure 4. However, by inspection, the images in Figures 4 and 5
are probably registered to within $\sim 1$ mas. In the context of our model
the protostar would lie at about the position $(-5, +1)$ mas in both
images.
The position uncertainties are all smaller than the sizes of the
plotted symbols.  Color corresponds to line-of-sight velocity, and the areas
of the plotted symbols are proportional to flux density.  We plot the
color bar from Figure \ref{fig:siomasrs} beneath the color bar
for this image to permit ready comparison.  
The inset shows the LSR velocity of the SiO maser spots
versus east-west offset.
The dashed line denotes a slope of 1 km s$^{-1}$ mas$^{-1}$.
\label{fig:siomasrs-vlba}}
\end{figure}

The two-armed linear structure of the SiO maser is 
suggestive of a diverging outflow on scales $\lesssim 25$ AU 
(Figure \ref{fig:siomasrs-vlba}). 
We propose that the SiO maser
emission may trace the limbs of a rotating conical bipolar outflow from
a massive protostar.  
In the context of this model, we can localize the relative position
of the exciting protostar on very small scales.  Specifically, 
we speculate that the protostar lies within several AU 
($\sim 5$ mas) of the intersection
of the northern and southern limbs (although the absolute position is
known to only $\sim 50$ mas).

The velocity gradient of the SiO maser emission, present along the 
northern (blue) arm and the southern (red) arm 
(Figure \ref{fig:siomasrs-vlba}), implies acceleration
along the putative outflow.  This acceleration is also visible in
the VLA image (Figure \ref{fig:siomasrs}), which shows bluer
emission to the east and redder emission to the west, as expected.
However, the VLA and VLBA data both show some blue emission that is
inconsistent with the fitted acceleration of $\sim 1$ km s$^{-1}$ mas$^{-1}$.
We suggest that this emission may not be associated with the proposed
diverging outflow, but it may instead be associated with a separate 
dynamical component ({e.g.,} a disk or associated wind, as in
Hollenbach et al. 1994).

The outflow traced on small scales by SiO maser emission appears 
to be reflected on large scales by
the H$_2$O maser complexes ${\cal A}$ and ${\cal B}$ .  
The position angle on
the sky of the SiO outflow ($\sim 105^{\circ}$)
is consistent with the
angle of the outflow traced by the motions and positions
of clusters ${\cal A}$ and ${\cal B}$ (Figure \ref{fig:dc}).
The bow shock H$_2$O maser structure ${\cal B}$ 
subtends the same opening angle as the SiO 
maser emission ($\sim 25^{\circ}$; Figures \ref{fig:dc} and
\ref{fig:siomasrs-vlba}).  Although ${\cal A}$ does not show a clear bowshock
morphology (perhaps due to inhomogeneities or gradients in the
ambient medium), the angle subtended by the strongest H$_2$O maser emission
in ${\cal A}$ is consistent with the proposed outflow.
The H$_2$O maser emission also shows the same general trend in LOS
velocity as the SiO maser emission, with bluer emission to the east
and redder emission to the west.
Thus, it appears that the SiO maser emission traces the limbs 
of a collimated 
diverging conical flow close to the exciting source, while the
H$_2$O masers trace the shocks that form where the ballistic outflow runs into
ambient material. 
(We note that the ultracompact
HII region W51d has a bowshock morphology whose position angle is close to that
of the H$_2$O outflow and bowshock, which may be a coincidence or may
be indicative of a larger scale outflow from the Dominant center
region along the same position angle.)

We estimate the angle of the putative outflow with
respect to the plane of the sky, $\theta$, by comparing the proper motions and
line-of-sight velocities of the H$_2$O maser complexes ${\cal A}$ and
${\cal B}$: 
\begin{equation}
\theta = \tan^{-1} (v_{\rm LOS} /
v_{\rm prop}) \sim 4^{\circ},
\label{eq:1}
\end{equation} 
where $v_{\rm LOS} = 5$
km s$^{-1}$ is half the difference in LOS velocity between 
${\cal A}$ and ${\cal B}$ 
and $v_{\rm prop} = 80$ km s$^{-1}$ is half the speed at which 
${\cal A}$ and ${\cal B}$ are separating on the sky.
However, because of the
large dispersion in the LOS velocities of H$_2$O maser clusters 
${\cal A}$ and ${\cal B}$, this is only a rough estimate.

We may also estimate $\theta$ by combining
LOS velocity information from the SiO maser
emission with the proper motions of the H$_2$O maser emission.
For purposes of discussion, we assume a common acceleration.
On small scales the acceleration is given  by
$a_1 = v_{\rm SiO}^2/{2 d_{\rm SiO}} = (v_1^2 \cos \theta)/
(2 d_1 \sin^2 \theta)$. Here, $d_1 = 30$ AU (4 mas) is the measured
transverse distance from the intersection of the two arms to the
easternmost maser spot in Figure \ref{fig:siomasrs-vlba},
and $v_1 = 4$ km s$^{-1}$ is the measured difference in line-of-sight
velocity between these features.  On larger scales,
the acceleration is given by $a_2 = v_{\rm H2O}^2/{2 d_{\rm H2O}} = 
(v_2^2)/(2 d_2 \cos \theta)$, where $d_2=2130$ AU ($0\rlap{.}''6$)
and $v_2 = 80$
km s$^{-1}$ are the transverse separation and transverse velocity
difference of the SiO maser and H$_2$O maser complexes ${\cal A}$
and ${\cal B}$ (where we assume that the transverse velocity of the SiO
maser is effectively $0$).  
If $a_1=a_2$, then
the angle of the outflow is given by 
\begin{equation}
\theta = \tan^{-1} \left(\frac{v_1}{v_2}
\sqrt{\frac{d_2}{d_1}} \right) \sim 20^{\circ}.
\label{eq:outflow-ang}
\end{equation}
However, a constant acceleration over $\sim 2200$ AU is unlikely
to occur.  For a decelerating flow,
$a_1 > a_2$ and $\theta < 20^{\circ}$.  If we assume 
constant acceleration of the jet out to $d_1$ with no acceleration
thereafter, we obtain $\theta = \tan^{-1} (v_1/v_2) \sim 3^{\circ}$,
consistent with the value estimated from Equation \ref{eq:1}.
Thus, the outflow probably lies within 20$^{\circ}$ 
of the plane of the sky.   

For the simple case of constant acceleration, we estimate the
acceleration of the outflow to be $0.5$ km s$^{-1}$ yr$^{-1}$,
which gives a line-of-sight acceleration of $0.2$ km s$^{-1}$ yr$^{-1}$.
For a decelerating flow, the implied acceleration on the scale
of the SiO maser emission is larger.
This is consistent with a 0.4  km s$^{-1}$ yr$^{-1}$  
drift in the line-of-sight velocity of individual SiO maser spectral
features reported by Fuente et al. (1989).  
However, the velocities of spectral components measured by others after
1989 (Hasegawa et al. 1986; Morita et al. 1992; this work)
are not entirely consistent with the trend fit by Fuente et al.
(1989), and the fluxes of components vary by over an order
of magnitude from observation to observation.  Although
suggestive, we conclude that the estimated acceleration from
Fuente et al. should be treated with caution.

Using our estimate of the angle of the outflow with respect to
the plane of the sky, we can also make an order of magnitude
estimate of the mass loss
rate for the putative protostar.  Assuming that the outflow
consists primarily of molecular hydrogen with 
a density of $n_{H_2} \sim 10^9$ cm$^{-3}$
(reasonable for SiO maser excitation; Elitzur 1992), we find
\begin{equation}
\dot{M} \ge \frac{m_p n \pi h^2 v_1}{2 \sin \theta} \sim 10^{-6}
\: M_{\odot} \: {\rm yr}^{-1}.
\label{eq:mdot}
\end{equation}
Here, $m_p$ is a proton mass, $h = 21$ AU is the distance
between the two limbs of SiO maser emission
measured at the easternmost maser spot in Figure \ref{fig:siomasrs-vlba},
$v_1 = 4$ km s$^{-1}$ is defined above, and $\theta=20^{\circ}$
is the upper limit on the 
angle of the outflow with respect to the plane of the sky under
the assumption of constant acceleration.  
We note that our estimate
is consistent with those for other massive protostars,
$\sim 10^{-6}$ $M_{\odot}$ yr$^{-1}$ \citep[{e.g.,}][]{LADA85}.

We estimate the line-of-sight 
velocity of the exciting protostar by averaging the
velocities of the SiO maser features on the north and south limbs
of the outflow:
$v_{\ast}  \sim  47$ km s$^{-1}$.
This is significantly different than the ${\rm 60 \: km \: s^{-1}}$ 
systemic velocity of the dominant center region obtained from 
interferometric observations of thermal NH$_3$ emission
\citep{ZH95, HGD83}.
However, the linewidth of the thermal NH$_3$ emission is
${\rm \sim 20 \: km \: s^{-1}}$, and the angular resolution of
the NH$_3$ observations is $1''$.  Thus, it is possible that
the molecular core 
has several components (possibly marked by the other clusters of H$_2$O
maser emission in the Dominant Center region), 
which are moving at different LOS
velocities, and perhaps the protostar that is driving
the outflow discussed here is associated with one of these components.

We postulate that the conical outflow rotates (counterclockwise)
about the axis of outward flow in order to explain the $\sim 5$
km s$^{-1}$ difference
in LOS velocity between the north and south limbs of SiO maser 
emission seen in Figure \ref{fig:siomasrs-vlba}.  The ratio of transverse
to rotational energy is $(v_{\rm flow} / v_{\rm rot})^2 \sim
(80/2.5)^2 \sim 1000$-- {i.e.,} most of the kinetic energy is in 
the outward flow component.  
Rotating bipolar outflows have been theorized to occur 
due to ``X-winds'' where magnetocentrifugal effects drive
the outflow \citep{NAJITA95, SHU+94}, or due to the
interaction of a spherical protostellar wind with a time-dependent
infall from a parent molecular cloud \citep{WS98}.
The limbs are presumably a preferred site of maser emission because
the longer path lengths and the similarity of projected LOS velocities 
result in greater amplification.
The limbs of a conical bipolar outflow traced by SiO masers have also
been observed in Orion-KL, though without rotation \citep{GREENHILL+98}.

Instead of a conical flow, the two arms of SiO maser emission 
in W51 could also
indicate the presence of two protostars, possibly comprising
a close binary system, wherein the north and south arms 
trace two separate outflows. 
While this model can also explain our results, 
the coincidence of the opening angle of the SiO maser
emission with the angle subtended by the bow shock structure
leads us to favor the conical outflow model.

We also consider a precessing jet model.  Were the precessing
jet to behave like a rigid rod, then magnetic coupling to ionized ambient
material might account for the observed velocity structure of the
SiO maser emission.
However, it is more likely that such a jet would instead be ballistic,
in which case the precession cone of the jet would have no ``limbs'' 
where the observed path length is greatest.  
In this case, if the jet comprises knots or bullets of
material, visible emission would appear to ``fill'' the precession
cone, something we do not see.
A possible ``fix'' is if the precessing jet shocks SiO in a 
rotating ``cocoon'' of gas, exciting maser emission.  If the decay time
for emission of shocked gas is long compared to the precession time-scale
of the jet, this model would lead to observed SiO maser emission
identical to the rotating conical outflow model.  If this model
is correct, then it should be possible to detect periodic changes
in the brightness of the arms of SiO maser emission detected with
the VLBA.  The variations in brightness of the two arms should
be out of phase.

\section{Conclusions}
\label{conclns}
We have resolved the structure of the W51-IRS2 SiO 
maser source, and linked it to a protostellar outflow
associated with two long-known sites of intense H$_2$O maser
emission.
We propose that the masers trace an accelerating bipolar protostellar outflow
inclined $<20^{\circ}$ with respect to the plane of the sky, 
which may detectably rotate about the axis of
flow within tens of AU of the protostar.  
We estimate the position angle of the flow to be $\sim 105^{\circ}$ up to
4200 AU from the central star.  The proper motions of H$_2$O maser
clusters bracketing the SiO maser source indicate an outflow
velocity of $\sim 80$ km s$^{-1}$ along this position angle, 
and one of these clusters appears
to trace a bow shock that subtends an angle consistent with the
opening angle suggested by the two limbs of SiO maser emission.
We estimate the acceleration of the outflow to be 
$\sim 0.5$ km s$^{-1}$ yr$^{-1}$, which is
consistent with the 0.4 km s$^{-1}$ yr$^{-1}$
line-of-sight velocity drift measured by Fuente et al. (1989).

In the larger context, this bipolar flow lies in within a $\sim 10^4$ AU
cloud core hosting multiple centers of H$_2$O maser emission.
The outflow does not extend to the limits of the core or to the other
centers of H$_2$O maser emission.  We estimate the line-of-sight velocity of 
the protostar that is driving the bipolar outflow to be
$\sim 47$ km s$^{-1}$,
significantly different than the systemic velocity of the NH$_3$
core (60 km s$^{-1}$).
We suggest that there may be
several star-forming fragments within this core, perhaps marked
by the other centers of H$_2$O maser activity.

\epsscale{0.8}

\epsscale{0.3}

\epsscale{0.7}

\epsscale{0.7}

\epsscale{0.7}
\end{document}